\documentclass[aps,twocolumn,superscriptaddress]{revtex4}
\usepackage{epsfig}
\usepackage{times}
\usepackage{color}
\usepackage{amsfonts}

\begin{document}

\title{Lying on networks: The role of structure and topology in promoting honesty}

\author{Valerio Capraro}
\email{v.capraro@mdx.ac.uk}
\affiliation{Department of Economics, Middlesex University, The Burroughs, London NW4 4BT, U.K.}

\author{Matja{\v z} Perc}
\affiliation{Faculty of Natural Sciences and Mathematics, University of Maribor, Koro{\v s}ka cesta 160, 2000 Maribor, Slovenia}
\affiliation{Department of Medical Research, China Medical University Hospital, China Medical University, Taichung 404, Taiwan}
\affiliation{Complexity Science Hub Vienna, Josefst{\"a}dterstra{\ss}e 39, 1080 Vienna, Austria}

\author{Daniele Vilone}
\affiliation{Laboratory of Agent Based Social Simulation, Institute of Cognitive Science and Technology, National Research Council, Via Palestro 32, 00185 Rome, Italy}
\affiliation{Grupo Interdisciplinar de Sistemas Complejos, Departamento de Matem\'aticas, Universidad Carlos III de Madrid, 28911 Legan\'es, Spain}

\begin{abstract}
Lies can have a negating impact on governments, companies, and the society as a whole. Understanding the dynamics of lying is therefore of crucial importance across different fields of research. While lying has been studied before in well-mixed populations, it is a fact that real interactions are rarely well-mixed. Indeed, they are usually structured and thus best described by networks. Here we therefore use the Monte Carlo method to study the evolution of lying in the sender-receiver game in a one-parameter family of networks, systematically covering complete networks, small-world networks, and one-dimensional rings. We show that lies which benefit the sender at a cost to the receiver, the so-called black lies, are less likely to proliferate on networks than they do in well-mixed populations. Honesty is thus more likely to evolve, but only when the benefit for the sender is smaller than the cost for the receiver. Moreover, this effect is particularly strong in small-world networks, but less so in the one-dimensional ring. For lies that favor the receiver at a cost to the sender, the so-called altruistic white lies, we show that honesty is also more likely to evolve than it is in well-mixed populations. But contrary to black lies, this effect is more expressed in the one-dimensional ring, whereas in small-world networks it is present only when the cost to the sender is greater than the benefit for the receiver. Lastly, for lies that benefit both the sender and the receiver, the so-called Pareto white lies, we show that the network structure actually favors the evolution of lying, but this only when the benefit for the sender is slightly greater than the benefit for the receiver. In this case again the small-world topology acts as an amplifier of the effect, while other network topologies fail to do the same. In addition to these main results we discuss several other findings, which together show clearly that the structure of interactions and the overall topology of the network critically determine the dynamics of lying.
\end{abstract}

\maketitle

\section{Introduction}
The conflict between lying and truth-telling is at the core of any social or economic interaction with asymmetric information. Lying, while sometimes interpreted as a sign of intelligence in children \cite{alloway2015liar} and a relatively common occurrence in adults to get out of awkward situations, can be detrimental to people, governments, organizations, firms, and ultimately to human societies as a whole. The cost of tax evasions in the USA alone, for example, has been estimated at 100 billion per year~\cite{gravelle2010tax}. Lying negatively affects also close personal relationships, being associated with marital dissatisfaction and friendship dissolution \cite{argyle1984rules,hendrick1981self}. Thus not surprisingly, researchers have sought to understand factors that determine dishonest behavior for years~\cite{mazar2008dishonesty, ariely2012honest, gino2009contagion, gino2011unable, shalvi2011justified, shalvi2012honesty, shalvi2015self, verschuere2011ease, biziou2015does, capraro2017does, capraro2018gender, erat2012white, gneezy2018lying, capraro2019time, fischbacher2013lies}.

Here we advance this subject by using methods of statistical physics. Indeed, the past two decades have significantly expanded the scope of physics beyond its traditional boundaries. Various aspects of economics \cite{mantegna1999introduction} and social sciences \cite{castellano_rmp09, orsogna_plr15, pastor_rmp15, wang_z_pr16, perc2017statistical} have benefited from the Monte Carlo method \cite{stanley_71, binder_88} and the coming-of-age of network science \cite{estrada2012structure, boccaletti_pr14, kivela_jcn14, barabasi_16}. In particular the social dynamics in general \cite{castellano_rmp09}, as well as more specific aspects of modern human societies, such as crime~\cite{orsogna_plr15}, gossiping~\cite{giardini2016evolution}, epidemics \cite{pastor_rmp15}, vaccination~\cite{wang_z_pr16}, and cooperation \cite{perc2017statistical}, have all been successfully studied using methods of physics and the gist of the `physics approach' \cite{perc_srep19}, which is to rationally select the key components of a system until the latter is fit to describe the essence of the problem at hand. These preceding developments certainly invite physics research into the realm of other types of moral behaviors~\cite{capraro2018grand}, including lying and honesty~\cite{capraro2019evolution}.

Part of the success of the Monte Carlo method relies on the fact that it can be used in evolutionary game theory to simulate the strategic evolution of the nodes of a network. The nodes are occupied by players that interact through a strategic game and then, after the accumulation of payoffs, update their strategies through a suitable imitation, replication, or exploration rule. In fact, this method has proven extremely useful to study the evolution of cooperation on lattices and heterogenous networks in social dilemmas \cite{santos2005scale,pacheco_prl06,pusch_pre08,ohtsuki_prl07,lee_s_prl11,mathiesen_prl11}, as well as to study strategic fairness in the ultimatum game~\cite{szolnoki2012defense,kuperman2008effect,eguiluz2009critical,da2009statistical,deng2011coevolutionary,gao2011coevolutionary,szolnoki2012accuracy,deng2012network,iranzo2012empathy,miyaji2013evolution}. However, to the best of our knowledge, no previous work has explored the evolution of honesty and lying on networks.

Here we make a first step in this direction. As a relatively simple but complete mathematical model of lying, we use the sender-receiver game~\cite{gneezy2005deception,erat2012white}. As we will show in the Mathematical model section, this paradigm is fundamentally different from previous games studied using the Monte Carlo method, such as the prisoner's dilemma and the ultimatum game. At the same time, it is particularly suitable for the application of the Monte Carlo method, and it allows us to study the evolution of four different types of lies. Namely, black lies, altruistic white lies, Pareto white lies, and spiteful lies (the Mathematical model section has all the definitions). While we have previously studied the evolution of lying using the sender-receiver game in well-mixed populations~\cite{capraro2019evolution}, it remains an open question whether and to what degree the fact that our interactions are commonly structured rather than well-mixed impacts the results. The evolution of lying in well-mixed populations was found to be strongly dependent on the type of lie, and it also displayed complex character that precluded generalizations over different parameters of the game. Since human interactions are not random, as we are more likely to interact within our social circles, such as with family, friends, or within our workplace with colleagues and coworkers, it is thus important to go beyond well-mixed populations and to study the evolution of lying in networks.

To that effect, we study the evolution of lying in a large family of networks, known as LASW networks~\cite{vilone2011random}. LASW networks are a one-parameter family of networks spanning from the one-dimensional ring to the complete network, as follows. One starts from the one dimensional ring and then adds each of the missing edges with probability $p\in[0,1]$. Therefore, if $p=0$, one remains with the one-dimensional ring, while $p=1$ returns a complete graph. When $p$ varies from $0$ to $1$ one obtains a number of intermediate cases of great theoretical and practical interest, such as small-world networks~\cite{watts_dj_n98}, which are thought to underline several sociological phenomena~\cite{milgram_pt67,kochen1989small}.

The paper is structured as follows. The Mathematical model section contains all the definitions. These definitions are grouped in three subsections. In the sender-receiver game subsection we describe the definition of the sender-receiver game, in the LASW network section we describe in self-sufficient detail this family of networks, and in the Monte Carlo method section we described the details of the model how it is simulated. The Results section reports all the main results and findings, whereas in the Discussion section we compare these to the results in the existing literature, and we also point out avenues for future research.

\ 

\section{Mathematical model}

\subsection{The sender-receiver game}

To study the evolution of dishonesty, we use the sender-receiver game. Among the many decision problems and strategic games that behavioral scientists have introduced to study people's dishonesty~\cite{fischbacher2013lies,mazar2008dishonesty,smith1991honest}, the sender-receiver game is particularly suitable for the application of the Monte Carlo method, because it is a game with two players and (practically) two strategies~\cite{capraro2019evolution}. This game was initially introduced by Uri Gneezy in ~\cite{gneezy2005deception}. Here, we adopt a subsequent version developed by Erat and Gneezy~\cite{erat2012white}. There are two potential allocations of money between the sender and the receiver, Option A and Option B. Without loss of generality, we can normalize the payoffs such that $A=(0,0)$ and $B=(r,s)$, with $r,s\in[-1,1]$. The first component represents the payoff of the receiver; the second component that of the sender. The sender privately rolls a six-face dice and looks at the outcome. Then the sender chooses a message to send to the receiver among six possible messages: ``The outcome was $i$'', with $i\in\{1,2,3,4,5,6\}$. After receiving the message, the receiver guesses the true outcome of the roll of the dice. If the receiver guesses the true outcome, then Option A is implemented as a payment; if the receiver does not guess the true outcome of the roll of the dice, then Option B is implemented.

Therefore, the sender has essentially two strategies: he either sends a truthful message, or not. Similarly, the receiver has essentially two strategies: she either believes the message sent by the sender, or not. If the receiver believes the sender, she reports the same number as the one sent by the sender; otherwise, if the receiver does not believe the sender, she draws randomly a number from the remaining five numbers of the dice. Therefore, the sender-receiver game can be written using the following bi-matrix:

\begin{center}
\begin{tabular}{ |c|c|c| }
 \hline
  & B & N \\
 \hline
 T & $0,0$ & $s,r$ \\
 L & $s,r$ & $\frac{4}{5}s$ ,$\frac{4}{5}r$  \\
 \hline
\end{tabular}
\end{center}
where $T$ stands for ``Truth'', $L$ stands for ``Lie'', $B$ stands for ``Believe'', and $N$ stands for ``Not Believing''. The ratios $\frac{4}{5}$ descend from the fact that, when the sender lies and the receiver does not believe the sender's message, then the receiver does not guess the true outcome of the roll of the dice with probability $\frac{4}{5}$. Therefore, these ratios directly descend from the formulation of the sender-receiver game proposed by Erat and Gneezy \cite{erat2012white}. In the discussion section, we will mention potential extensions to be explored in future research.

Another positive sides of the sender-receiver game, compared to other measures of dishonesty, is that it allows to study different types of lie. Following the taxonomy introduced by Erat and Gneezy~\cite{erat2012white}, we introduce four types of lies, depending on whether lying harms or benefits the agents:
\begin{itemize}
    \item Pareto white lies are those that benefit both the sender and the receiver: $r,s>0$.
    \item Altruistic white lies are those that benefit the receiver and harm the sender: $r>0$, $s<0$.
    \item Black lies are those that benefit the sender and harm the receiver: $r<0$, $s>0$.
    \item Spiteful lies are those that harm both the sender and the receiver: $r,s < 0$.
\end{itemize}

\subsubsection{Equilibrium analysis}

The distinction among different types of lie turns out to be useful for the equilibrium analysis. In order to find the equilibria of the game, it is convenient to describe it as a symmetrical two-player game. Since, in our model, each player acts as both sender and receiver, we have in practice four composed pure strategies $\sigma\equiv(X_s,X_r)$, where $X_s\in\{T,L\}$ and $X_r\in\{B,N\}$ are the basic strategies as a sender and receiver, respectively. As we are going to illustrate in Sec.~\ref{mc_sec}, we have a population undergoing an evolutionary dynamics, which implies that the Nash equilibria can be found by means of the replicator equations~\cite{hofbauer_98}:

\begin{equation}
\left\{
\begin{array}{l}
\dot{x}_{\sigma} =x_{\sigma}\cdot(\Pi_\sigma-\bar{\Pi}) \\
\ \\
\sum_\sigma x_\sigma =1 \ ,
\end{array}
\right.
    \label{repl_eq}
\end{equation}

\noindent where $x_\sigma$ and $\Pi_\sigma$ are the densities and the average payoffs of each strategy $\sigma$, respectively, and $\bar{\Pi}$ is the average global payoff. The normalization condition in Eq.~(\ref{repl_eq}) represents the conservation of the population's size. After some laborious but easy calculation the quantities $\Pi_\sigma$ and $\bar\Pi$ can be computed, and by imposing the stationary condition on Eqs.~(\ref{repl_eq}) it is straightforward to find the equilibria.

In case of spiteful lies, there are two equilibria in pure strategies (that is, configurations where all the agents share the same strategy) $(T,B)$ and $(L,N)$,
and one mixed equilibrium $x_{(T,B)}= 1/6,\ x_{(L,N)}=5/6$. In case of altruistic or black lies, there is only one equilibrium in mixed strategies, which is, again, $x_{(T,B)}= 1/6,\ x_{(L,N)}=5/6$. Finally, in case of Pareto white lies, there are two equilibria in pure strategies, $(T,N)$, $(L,B)$, and one equilibrium in mixed strategies, which is, once again, $x_{(T,B)}= 1/6,\ x_{(L,N)}=5/6$. The cases $r=0$ and/or $s=0$ are straightforward, because the corresponding agents are indifferent between the two available strategies.

This analysis also highlights the fundamental difference between the sender-receiver game and the previous games that have been studied using the Monte Carlo method, the prisoner's dilemma and the ultimatum game, which have radically different set of equilibria.

\subsection{LASW networks}

A LASW network is defined starting from a regular one-dimensional ring with $N$ nodes, each one connected to its $2m$ ($m\in\mathbb{N}$) nearest neighbours, so that there are $mN$ links (or edges). The topology is then modified by adding new links, that is, we add each of the $[N(N-1)/2-mN]$ initially missing edges with probability $p$. Therefore, by varying $p$ we can tune the topology from euclidean to the complete-network as desired~\cite{vilone2011random}. In particular, we have regular, one-dimensional lattice for $p\rightarrow0^+$, the Watts-Strogatz (WS) small-world topology~\cite{watts_dj_n98} for $0<p\lesssim p^*$, a random network (RN)~\cite{erdos_pmd59} for $p^*\lesssim p<1$, whilst for $p\rightarrow1^-$ we get a complete graph (that is, we recover the mean-field configuration); the critical link adding probability is $p^*=2m/N$~\cite{vilone2011random}.

The main topological differences among these four regimes can be described by the behaviour of the {\it diameter} $D$ and the {\it clustering coefficient} $\chi$ of the network. $D$ is defined as the mean shortest path between two random nodes; $\chi$ is defined as the probability that two nodes that share a neighbour are also neighbour themselves. A ring has high diameter (with the exception of the case $m=1$) and high clustering coefficient; a WS network is instead characterized by small diameter and high clustering coefficient; in random networks both quantities are small; finally, in a complete graph, one has $D=1$ and $\chi=1$. 

\subsection{The Monte Carlo method}
\label{mc_sec}

We consider the sender-receiver game among $N$ agents, who interact pairwise in a LASW network. At each round of the game, one agent plays in the role of the sender, and the other agent plays in the role of the receiver. The role of an agent is randomly determined at the beginning of the interaction. When playing as a sender, an agent can either tell the truth ($T$) or lie ($L$); when playing as a receiver, an agent can either believe ($B$) the message sent by the sender, or not ($N$). This results in four different strategies: $(T,B)$, $(T,N)$, $(L,B)$, and $(L,N)$. Initially, each agent is randomly assigned to either $T$ or $L$ (when she plays as a sender), and to either $B$ or $N$ (when she plays as a receiver).

We simulate the game using the Monte Carlo method. The following elementary steps apply. First, an agent $x$ is randomly drawn from the population. Agent $x$ then plays the sender-receiver game with four randomly chosen neighbours in a pairwise manner as described above, thereby obtaining the payoff $\pi_x$. Note that, for $p=0$, each agent has exactly four neighbours, therefore, in this case, there is no random selection, and $x$ plays with all his neighbours; when $p$ increases, the number of neighbours statistically increases, and therefore in this case, there is an actual random selection of the four neighbours with whom $x$ plays. Then, another agent $y$ is randomly drawn from the population, and he also plays the sender-receiver game with four randomly selected neighbours, thereby obtaining the payoff $\pi_y$. Finally, agent $y$ imitates the strategy of agent $x$ with the probability $w=\{1+\exp[(\pi_{y}-\pi_{x})/K]\}^{-1}$, where $K$ encapsulates the uncertainty during the strategy adoption process. When $K \to \infty$, payoffs lose importance and strategies change at random; conversely, when $K \to 0$, agent $y$ imitates $x$ only if $\pi_{x} > \pi_{y}$; between these two limits, the strategies of better performing agents tend to be imitated, although under-performing strategies are imitated as well, with non-zero probability. In reality, this may be due to errors in the decision making, imperfect information, and external influences that may adversely affect the computation of the other player's payoff. Without loss of generality, we set $K=0.1$, in agreement with previous work showing this to be a representative value~\cite{perc2017statistical}.

\begin{figure*}
  \centering
  \includegraphics[width=157mm]{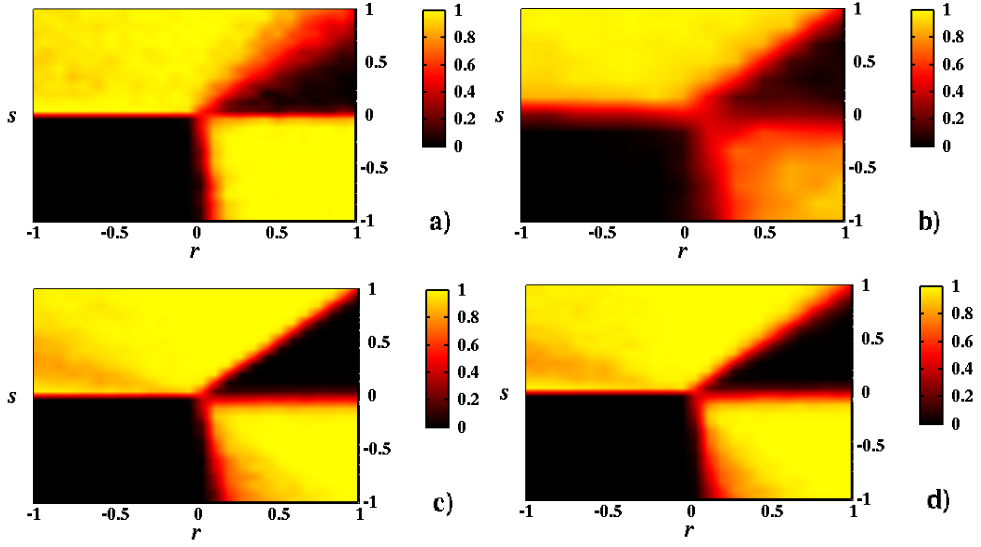}
  \caption{Final density of liars in {\bf a)} the well-mixed network, {\bf b)} the one-dimensional ring, and two small-world networks with {\bf c)} $p=0.00025$ and {\bf d)} $p=0.001$. Systems of size $N=500$. Averages over 300 independent realizations.}
  \label{liar-comp}
\end{figure*}

The time is measured in Monte Carlo steps (MCS), whereby one MCS corresponds to executing all three elementary steps $N$ times. During one MCS, each agent changes strategy, on average, once. For a systematic numerical analysis, we have determined the fraction of strategies in the stationary state when varying the values of $s$ and $r$. In order to obtain adequate accuracy, we have used large system sizes, varying from $N=500$ to $1000$, as well as long enough thermalization and sampling times, varying from $10^4$ to $10^6$ MCS. To further remove statistical fluctuations, we have averaged the final outcome over up to $2000$ independent realizations.

\ 

\section{Results}

\subsection{Final densities of liars and believers across lie type and networks}

As a first step of our analysis, we look at the final densities of liars and believers, as a function of lie type (parameters $r$ and $s$) and network (parameter $p$). In this and in the following analyses, we focus on four prototypical values of $p$: $p=0$ (ring), $p=0.00025$ $p=0.001$ (two small-world networks), and $p=1$ (well-mixed population). We conducted numerical simulations also with several other $p$ values, but the results do not qualitatively differ from the above cases. Specifically, when $p<0.1$ (small-world networks), the pattern of results is qualitatively very similar to the cases $p=.00025$ and $p=.001$; in case of values of $p$ greater than $0.1$ (random networks), numerical simulations show that the differences with the complete network become tinier and tinier, therefore we report directly the limit, well-mixed, case, $p=1$.

\begin{figure*}
  \centering
  \includegraphics[width=157mm]{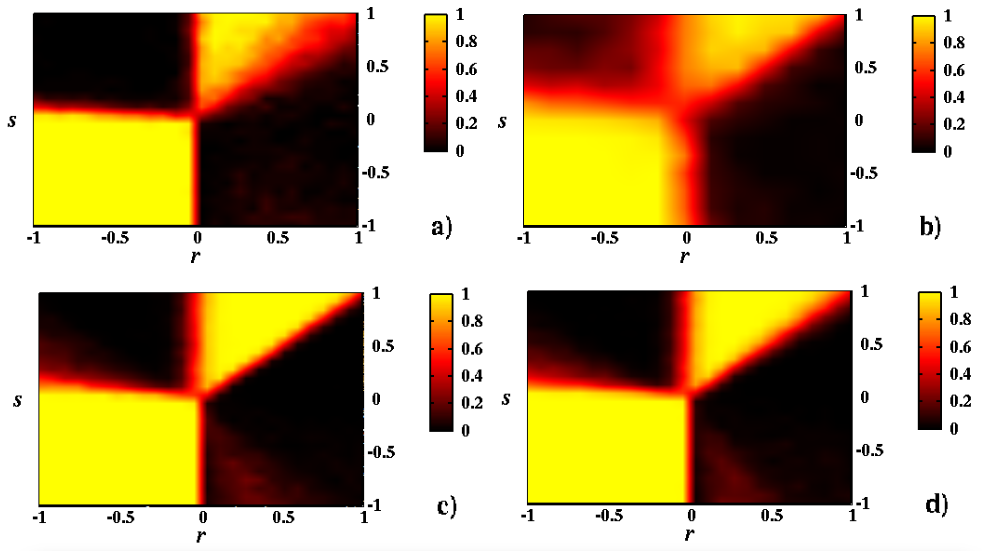}
  \caption{Final density of believers in {\bf a)} the well-mixed network, {\bf b)} the one-dimensional ring, and two small-world networks with {\bf c)} $p=0.00025$ and {\bf d)} $p=0.001$. Systems of size $N=500$. Averages over 300 independent realizations.}
  \label{belv-comp}
\end{figure*}

We start with the liars. Figure~\ref{liar-comp} highlights several differences between the final densities of liars in small-world networks (lower left panel: $p=0.00025$; lower right panel: $p=0.001$) and the final densities of liars in well-mixed populations (upper left panel). Specifically, both in the domain of black lies ($r<0$, $s>0$) and in the domain of altruistic white lies ($r>0$, $s<0$), the evolution of lying is disfavored in the two small-world networks, compared to the complete network, but only below the diagonal $r=-s$. By contrast, in the domain of Pareto white lies ($r,s>0$), the differences between the small-world networks and well-mixed populations tend to be concentrated right above the diagonal $r=s$; in particular, lying is favoured in small-world networks compared to the complete network. The case of spiteful lies is trivial: honesty evolves with frequency 1 independently of the network.

Moving to the case of the one-dimensional ring ($p\to0^+$, upper right panel). In this case, we note that the most evident differences, compared to the well-mixed population, appear in the domain of altruistic white lies ($r>0$, $s<0$) and, to a lesser extent, in the domain of Pareto white lies ($r,s>0$) and in the domain of black lies ($r<0$, $s>0$). Specifically, compared to the well-mixed population, altruistic white lies are less frequent than they are in the one-dimensional ring. In the domain of black lies, lying appears to evolve with very similar frequency in the two networks. By contrast, Pareto white lies are more frequent in the one-dimensional ring, compared to the well-mixed population, but this happens only just above the diagonal $r=s$; for other values of $r$ and $s$ in the same quadrant, the two networks behave roughly the same. Finally, in the domain of spiteful lies ($r,s<0$) the two networks behave identically.

Coming to the evolution of believers, Figure~\ref{belv-comp} highlights that the evolution of believers mirror the evolution of liars discussed above, although with some differences. For $p=0.00025$ (lower left panel) and $p=0.001$ (lower right panel) the differences between the two small-world networks and the complete network are concentrated in the domains of altruistic white lies, black lies, and Pareto white lies. In the first two domains, the differences are concentrated in the region in which $s$ is very small, where believing the sender's message is more likely to evolve. In the third domain, the differences are concentrated right above the diagonal $r=s$, where, again, believing the sender's message is more likely to evolve, compared to the well-mixed case.

As in the case of liars, the one-dimensional ring ($p=0$, upper right panel), gives rise to slightly different results. Specifically, in the domain of black lies, the evolution of believers is favored in the one-dimensional ring, compared to the well-mixed case, especially for values of $s$ close to $0$ or values of $r$ close to $1$. In the case of Pareto white lies, instead, the evolution of believers somewhat reflects the evolution of liars observed above: the evolution of believers is favored in the one-dimensional case, compared to the well-mixed case, but only just above the diagonal $r=s$; in the other parts of the Pareto white lies quadrant, there are no major differences between the two networks.

\begin{figure*}
  \centering
  \includegraphics[width=137mm]{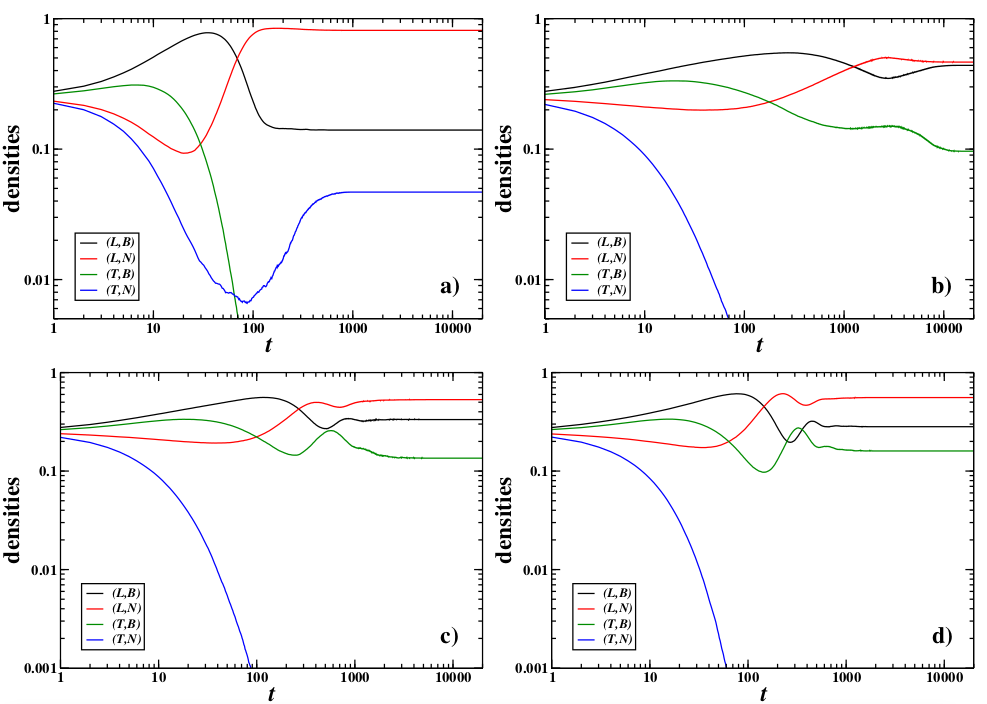}
  \caption{Time evolution of the four pure strategy profiles $(T,B)$, $(L,B)$, $(T,N)$, and $(L,B)$ for $s=0.2,\ r=-1$ for {\bf a)} well-mixed, {\bf b)} one-dimensional ring, and two small-world networks, {\bf c)} $p=0.00025$ and {\bf d)}  $p=0.001$. System size $N=500$, average over 3000 independent realizations, random initial conditions.}
  \label{strategies-comp3}
\end{figure*}

\subsection{Time evolution of liars and believers across networks}

\subsubsection{Black lies}

In the previous section, we have shown that the network structure affects the final densities of liars and believers. The previous section, however, does not allow to answer the question of \emph{which} types of liars and believers are more or less likely to evolve in small-world networks and the one-dimensional ring, compared to the complete network. Indeed, there are two types of liars: those who, when acting as receivers, believe the sender's message; and those who, when acting as receivers, do not believe the sender's message. Similarly, there are two types of believers: those who, when acting as senders, tell the truth; and those who, when acting as sender, lie. Therefore, to have a better understanding of what type of agents evolve, here we analyze, across the four networks under consideration, the time evolution of each of the four pure strategy profiles $(T,B)$, $(L,B)$, $(T,N)$, and $(L,B)$. For each lie type, we study one pair $(s,r)$. To select this pair, we follow a pragmatic approach. We start from black lies.

Figure~\ref{liar-comp} and Figure~\ref{belv-comp} suggest that, in the domain of black lies ($r<0, s>0$), the differences are concentrated below the diagonal $r=-s$, and, in particular, for $s$ small and $r$ close to $-1$. Therefore, we select $r=-1$ and $s=-0.2$.

Figure~\ref{strategies-comp3} reports the time evolution of the four pure strategy profiles for these values of $r$ and $s$. Comparing the upper left panel (well-mixed population) with the other panels, we find a number of differences. The strategy profile $(T,N)$ (blue line), which survives in well-mixed populations, quickly vanishes in all other networks. The strategy $(L,B)$ (black line) survives in all networks, although with different frequencies: in the well-mixed population it survives with frequency around $0.15$, while, in the other networks, the final density more than doubles. The strategy $(L,N)$ (red line) too survives in all networks, although with different frequencies: in the well-mixed population it survives with frequency $0.8$, while in all other networks the final density is around $0.5$. But the most interesting case is the case of the strategy profile $(T,B)$ (green line). This strategy profile quickly vanishes in well-mixed populations, but it survives with a relatively high frequency (around $0.1$) in all other networks. Importantly, the vanish of $(T,N)$ is more than compensated by the emergence of $(T,B)$. Indeed, in well-mixed populations, $(T,N)$ survives with frequency less than $0.05$, while, in the other networks, $(T,B)$ survives with frequency around $0.1$ and, in some cases (e.g., $p=0.001$) even close to $0.2$ Therefore compared to well-mixed populations, small-world networks and the one-dimensional ring have a net positive effect on honesty among senders. A similar argument holds for believers: the frequencies of both the strategy profile $(T,B)$ and $(L,B)$ are greater in the small-world networks and the one-dimensional ring, compared to the complete network. Therefore, there is a net overall positive effect on believing among receivers, although this net effect is stronger in the small-world networks compared to the one-dimensional ring.

\begin{figure*}
  \centering
  \includegraphics[width=137mm]{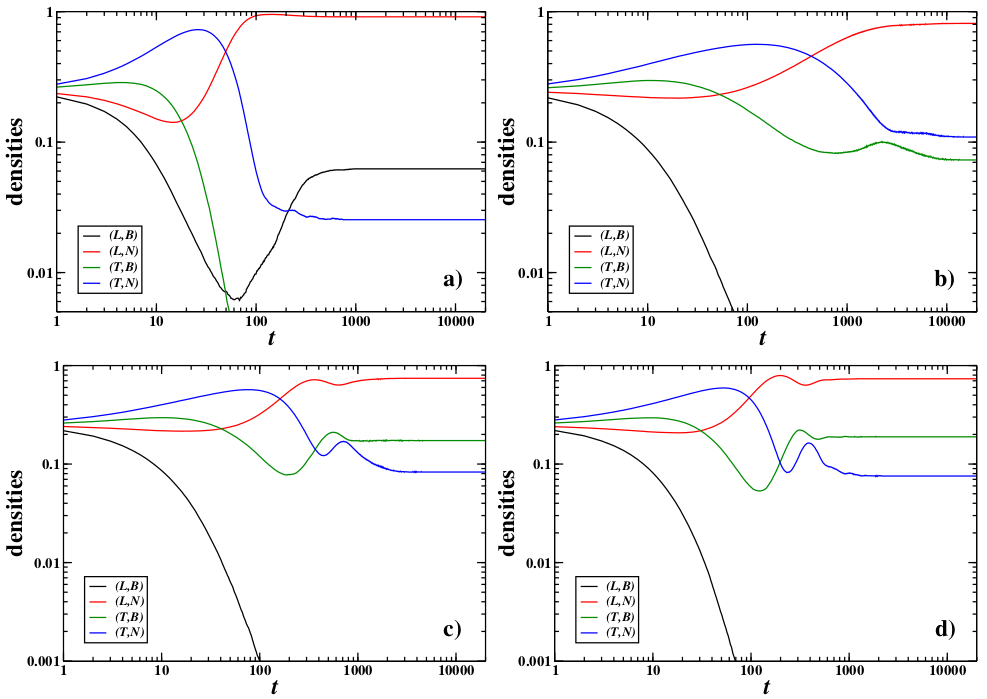}
  \caption{Time evolution of the four pure strategy profiles $(T,B)$, $(L,B)$, $(T,N)$, and $(L,B)$ for $s=-1,\ r=0.3$ for {\bf a)} well-mixed, {\bf b)} one-dimensional ring, and two small-world networks, {\bf c)} $p=0.00025$ and {\bf d)}  $p=0.001$.. System size $N=500$, average over 3000 independent realizations, random initial conditions.}
  \label{strategies-comp}
\end{figure*}

\subsubsection{Altruistic white lies}

Next, we study the time evolution of the four pure strategy profiles in a case in which lying benefits the receiver at a cost to the sender. Specifically, since Figure~\ref{liar-comp} and Figure~\ref{belv-comp} suggest that the differences between the networks are concentrated below the diagonal $r=-s$, we opted for illustrating the time evolution of the four pure strategy profiles using the parameters $r=0.3$ and $s=-1$.

Figure~\ref{strategies-comp} reports the outcomes of our numerical simulations. Compared to the upper left panel (well-mixed population), we find several differences. One clear difference concerns the evolution of $(T,B)$: while this strategy profile quickly vanishes in well-mixed populations, it survives in all other cases. In particular, in the two small-world networks, it survives with a density around $0.2$, while in the case of the one-dimensional ring, it survives with frequency around $0.08$. A similar pattern, although slightly less evident, emerges in case of the strategy profile $(T,N)$: in well-mixed populations, it survives with frequency less than $0.03$; in small-world networks it survives with frequency around $0.08$; in case of the one-dimensional ring, the final density is even above $0.1$. Clearly, the emergence of $(T,B)$ and $(T,N)$ when passing from complete to non-complete networks comes at the price of the other two strategy profiles, $(L,B)$ and $(L,N)$, both of which are less likely to evolve. Particularly evident is the case of $(L,B)$, which survives with non-zero frequency in well-mixed populations, but it quickly vanishes in all other cases. In sum, as in the black lie case discussed in the previous section, also in this case the presence of a non-complete network favours the evolution of honesty among senders. By contrast, the evolution of believing among receivers is favoured, compared to the well-mixed case, only in the two small-world networks, where the increase in the frequency of $(T,B)$ (equal to 0.2) more than counterbalance the decrease in the frequency of $(L,B)$ (equal to 0.06); in the one-dimensional ring, the increase in the frequency of $(T,B)$ (about 0.07) is very similar to the decrease in the frequency of $(L,B)$ (equal to 0.06).

\subsubsection{Pareto white lies}

Then, we study the time evolution of the four pure strategy profiles in a case in which lying benefits both the sender and the receiver. Specifically, since Figure~\ref{liar-comp} and Figure~\ref{belv-comp} suggest that the differences across networks are located slightly above the diagonal $r=s$, we opted for illustrating the evolution for $r=0.4$ and $s=0.6$.

\begin{figure*}
  \centering
  \includegraphics[width=137mm]{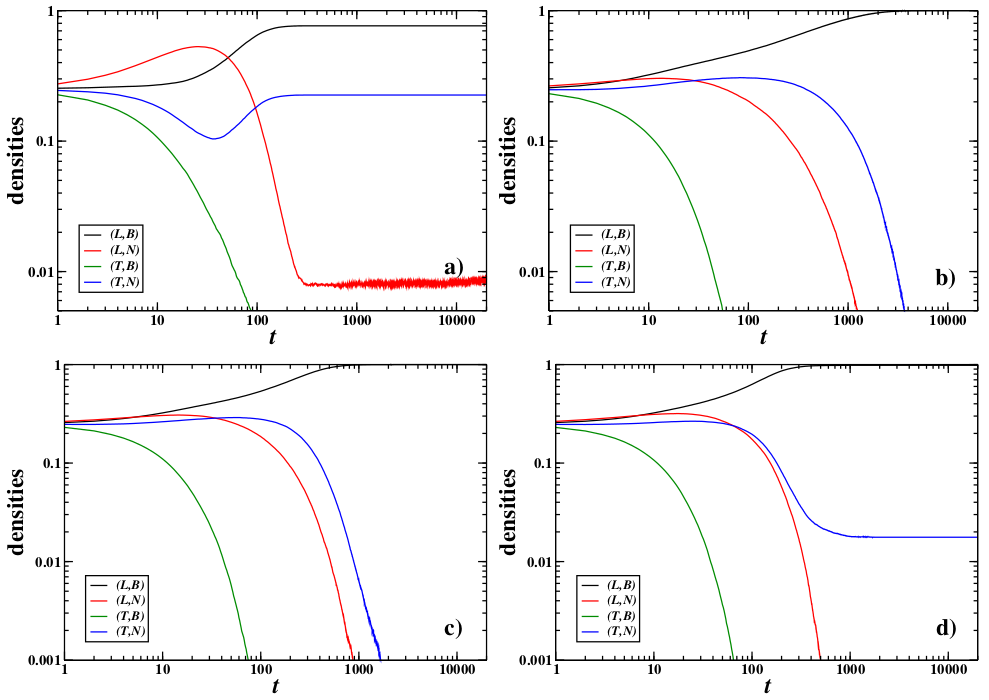}
  \caption{Time evolution of the four pure strategy profiles $(T,B)$, $(L,B)$, $(T,N)$ for $s=0.6$ and $r=0.4$ for {\bf a)} well-mixed, {\bf b)} one-dimensional ring, and two small-world networks, {\bf c)} $p=0.00025$ and {\bf d)}  $p=0.001$.. System size $N=500$, average over 3000 independent realizations, random initial conditions.}
  \label{strategies-comp2}
\end{figure*}

Comparing the upper left panel of Figure~\ref{strategies-comp2} (well-mixed population) with the other panels, we note an important difference. In well-mixed population, at the steady state, there is a co-existence of three pure strategy profiles, $(L,B)$ with frequency slightly below 0.8, $(T,N)$, with frequency slightly below $0.2$, and $(L,N)$ with a small frequency around 0.09. By contrast, in small-world networks and in the one-dimensional ring, only the strategy profile $(L,B)$ survives, while all other strategy profiles quickly vanish, apart from the case $p=0.001$, where there is a small residual of the pure strategy profile $(T,N)$, which survives with density below $0.02$. In sum, compared to the complete network, small-world networks and the one-dimensional ring have the effect of favouring the evolution of lying among senders and believing among receivers.

\subsubsection{Spiteful lies}

\begin{figure*}
  \centering
  \includegraphics[width=137mm]{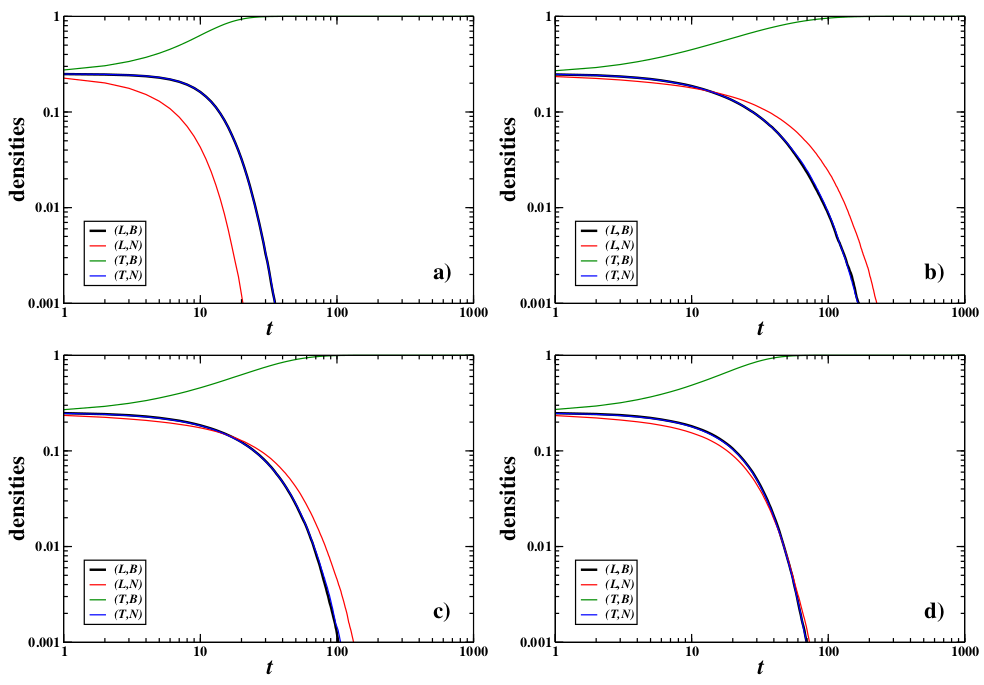}
  \caption{Time evolution of the four pure strategy profiles $(T,B)$, $(L,B)$, $(T,N)$, and $(L,B)$ for $s=-0.5$ and $r=-0.5$ for {\bf a)} well-mixed, {\bf b)} one-dimensional ring, and two small-world networks, {\bf c)} $p=0.00025$ and {\bf d)}  $p=0.001$.. System size $N=500$, average over 3000 independent realizations, random initial conditions.}
  \label{strategies-comp4}
\end{figure*}

We finally report the time evolution of the four strategy profile in the case of spiteful lies. Figure~\ref{liar-comp} and Figure~\ref{belv-comp} suggest that the evolution does not depend on the network, and that senders quickly learn that their best strategy is to tell the truth, while receivers quickly learn that their best strategy is to believe the sender's message. Our simulations confirm this finding. In Figure~\ref{strategies-comp4} we report the results of the simulations for $r=s=-0.5$. As expected, only the strategy profile $(T,B)$ survives, while all others quickly vanish.

\subsection{Steady state spatial configuration}

We have also conducted a set of simulations to explore the spatial configuration of the steady state. Our aim was indeed to explore whether certain strategy profiles tended to cluster together. However, interestingly, all the simulations that we have conducted converged to a unique strategy profile. Therefore, the final densities discussed above should be interpreted as the probability that a single realization evolves in such a way that all agents end up playing a given strategy profile.

\ 

\section{Discussion}

We have used the Monte Carlo method to study the evolution of lying in a set of networks including small-world networks and the one-dimensional ring. As a measure of dishonesty, we have used the sender-receiver game~\cite{gneezy2005deception,erat2012white}, a game that is fundamentally different from those that have been used in previous research applying the Monte Carlo method to evolutionary game theory on networks, such as the prisoner's dilemma and the ultimatum game. Our research shows that the spatial structure has a non-trivial effect on the evolution of the strategies, which depends significantly on the consequences of lying and telling the truth. The only trivial case is when lying harms both the sender and the receiver, i.e., when we have spiteful lies. In this case, regardless of the network, senders quickly learn that their best strategy is to tell the truth and receivers quickly learn that their best strategy is to believe the sender's message. In the case of black lies, that is those that harm the receiver at a cost for the sender, we found major differences across networks located below the diagonal $r=-s$. In this domain, Monte Carlo simulations show that honesty is more likely to evolve in small-world networks and, to a lesser extent, in the one-dimensional ring, compared to the well-mixed case. A slightly different result holds for receivers, where the differences are concentrated in a region in which $s$ is very small. Here, believing the sender's message is more likely to evolve in the one-dimensional ring and, to a lesser extent, in the small-world network, compared to the well-mixed case. In the case of altruistic white lies, that is those that benefit the receiver at a cost to the sender, we find that the regions in which the differences between the networks are located depend on the network topology. In the one-dimensional ring, we find that honesty is more likely to evolve, compared to the well-mixed case, regardless of the specific payoffs $r$ and $s$. By contrast, in small-world networks, honesty is more likely to evolve, compared to the well-mixed case, but only below the $r=-s$ diagonal. Slightly different are the results for receivers. In this case, the evolution of believing in the one-dimensional ring is identical to that in the well-mixed population. By contrast, the evolution of believing is favored in small-world networks, compared to well-mixed populations, but only below the $r=-s$ diagonal. Finally, in the case of Pareto white lies, that is those that benefit both the sender and the receiver, the major differences across networks are slightly above the $r=s$ diagonal, where both lying and believing are more likely to evolve compared to the well-mixed case. In sum, our findings show that the spatial structure has a highly non-trivial effect on the evolution of honesty and lying in the sender-receiver game. We stress that, with the exception of spiteful lies, our analysis shows that the final densities largely depend on the specific parameters, $r$ and $s$, even within a given lie type. This implies that the steady states almost never coincide with the equilibria of the game. The only case in which this happens is in the domain of spiteful lies, where the time evolution quickly converges to one of the equilibria, the pure strategy profile $(T,B)$, while discarding the other two equilibria, $(L,N)$ and the mixed one $x_{(T,B)}= 1/6,\ x_{(L,N)}=5/6$. The reason why these latter two equilibria are discarded in favor of the former one is because $(T,B)$ maximizes the payoff of both players, and therefore it is more likely to be imitated.

Most previous work on the evolution of lying used games different from the sender-receiver game. For example, a stream of research used the Philip Sidney game~\cite{smith1991honest,grafen1990biological,szamado2011cost,catteeuw2014evolution}, where the sender is with some probability initially either healthy, or with the remaining probability needy. The sender can then either pay a cost to signal his state to the receiver or do nothing. If the sender decides to signal his state, he can lie about it. Indeed, the receiver does not know the state of the sender, but can only observe the signal. After observing the signal if the latter is sent, the receiver decides whether to donate his resource to the sender. The sender-receiver game used here departs from the Philip-Sidney game along two dimensions. Firstly, in the sender-receiver game the signalling is cost-free. Even in this case, our results demonstrate that honesty can evolve in some circumstances, even when lying is self-serving (black lies). Secondly, the sender-receiver game differs from the Philip Sidney game in that it allows to study the evolution of lying not only in the domain of black lies, but also in the domains of spiteful lies, Pareto white lies, and altruistic white lies. Therefore, the sender-receiver game mathematically describes a much broader class of lies, and accordingly richer are the insights it affords. Another stream of previous research focused on cooperation games preceded by a commitment phase in which agents can make promises about what they will do in the subsequent cooperation game~\cite{han2013good,han2015avoiding,pereira2017evolution}. Our approach differs from this line of work in that we focus on honesty and believing, with no consequences on subsequent games. This allows us to clearly identify the four types of lies, and to study the evolution of honesty as a function of the type of lie.

As any research, also ours has some limitations. In the first place, we studied the evolution of lying only on a specific family of networks. This one-parameter family allows us to continuously move from complete networks to small-world networks and further to the one-dimensional ring. Therefore, this left out a number of other networks that are thought to emerge in many social settings, such as scale-free networks~\cite{barabasi_s99} or echo chambers \cite{williams2015network}. Previous work has explored the evolution of cooperation in these networks \cite{santos2005scale,santos2006evolutionary,santos2008social,evans2018opinion}. Future works should therefore investigate the evolution of lying in these networks, as well as others, such as interdependent and multilayer networks \cite{boccaletti_pr14, kivela_jcn14, wang_z_epjb15}. Secondly, we studied the evolution of lying in a somewhat natural condition, in which there is no punishment or reward. These mechanisms are well-known to favor cooperative behavior~\cite{boyd_s10,boyd1992punishment,fehr_aer00,gurerk_s06, panchanathan_n04,milinski_pnas06,andreoni_aer03,szolnoki_epl10,szolnoki_prx13,capraro2016partner, szolnoki_prx17, fang_prsa19}, and it is likely that they might also favor honest behavior along similar mechanisms reported for cooperation. Accordingly, future works should explore mechanisms to promote the evolution of honesty, starting with punishment and rewarding, as well as by other means, such as reputation or shame, which are by experience often fit to work in reality. Thirdly, we studied the evolution of lying in situations in which the players are forced to play every round of the interaction. In reality, it sometimes happens that players are unable to participate in an interaction, due to unforeseen circumstances, or simply because they decide to opt out from one particular interaction. Previous research has explored the effect of opting out on the evolution of cooperation \cite{ginsberg2019evolution}. Future research should extend this line of work in the context of lying. Finally, we studied the evolution of lying in the sender-receiver game as proposed by Erat and Gneezy \cite{erat2012white}. In this game, the set of potential private pieces of information available to the sender has cardinality six, and this ultimately generates the coefficient $\frac{4}{5}$ in the bi-matrix representing the game. Of course, one could generalize the sender-receiver game to information sets of any (finite) cardinality and study the evolution of lying as a function of a new parameter, representing the cardinality of the information set. Future work should explore the evolution of lying in this generalization of the sender-receiver game.

\ 

\begin{acknowledgments}
Matja{\v z} Perc was supported by the Slovenian Research Agency (Grant Nos. J4-9302, J1-9112, and P1-0403). Daniele Vilone was supported by the European Union's Horizon 2020 Project PROTON (Grant No. 699824).
\end{acknowledgments}

\end{document}